\documentclass[twocolumn,preprintnumbers,amsmath,amssymb]{revtex4}
\usepackage{epsfig}
\begin{document}

\title{Ferromagnetic resonance study of polycrystalline Fe$_{1-x}$V$_x$ alloy thin films}
\author{J-M. L. Beaujour, A. D. Kent} \affiliation{Department of Physics, New York University, 4
Washington Place, New York, NY 10003, USA}
\author{J. Z. Sun}
\affiliation{IBM T. J. Watson Research Center, Yorktown Heights,
NY 10598, USA}
\date{\today}

\begin{abstract}
Ferromagnetic resonance has been used to study the magnetic properties and magnetization dynamics of polycrystalline Fe$_{1-x}$V$_{x}$ alloy films with $0\leq x < 0.7$. Films were produced by co-sputtering from separate Fe and V targets, leading to a composition gradient across a Si substrate. FMR studies were conducted at room temperature with a broadband coplanar waveguide at frequencies up to 50 GHz using the flip-chip method. The effective demagnetization field $4  \pi M_{\mathrm{eff}}$ and the Gilbert damping parameter $\alpha$ have been determined as a function of V concentration. The results are compared to those of epitaxial FeV films.
\end{abstract}
\maketitle
\section{Introduction}
A decade ago, it was predicted that a spin polarized current from a relatively thick ferromagnet (FM) could be used to switch the magnetization of a thin FM \cite{slonczewski}. Since then, this effect, known as spin-transfer, has been demonstrated in spin-valves \cite{kent1} and magnetic tunnel junctions \cite{tj}. In a macrospin model with collinear layer magnetizations, there is a threshold current density $J_{\mathrm{c}}$ for an instability necessary for current-induced magnetization switching of the thin FM layer \cite{slonczewski,jsun}:    
\begin{equation}
\label{e.00} J_{\mathrm{c}} = \frac{2 e \alpha M_{\mathrm{s}} t_{\mathrm{f}} (H_{\mathrm{k}}+2 \pi M_{\mathrm{s}})}{\hbar \eta},
\end{equation}
where $\alpha$ is the damping constant. $t_{\mathrm{f}}$ and $M_{\mathrm{s}}$ are the thickness and the magnetization density of the free layer, respectively. $H_\mathrm{k}$ is the in-plane uniaxial anisotropy field. $\eta$ is the current spin-polarization. In order for spin-transfer to be used in high density memory devices $J_{\mathrm{c}}$ must be reduced. From Eq. \ref{e.00} it is seen that this can be achieved by employing materials with low $M_{\mathrm{s}}$ and $\alpha$ in spin-transfer devices or, equivalently materials with low Gilbert damping coefficients,  $\mathrm{G}=\alpha M_{\mathrm{s}} (g \mu_{\mathrm{B}}/ \hbar)$.

Very recently, an experimental study of epitaxial FeV alloy thin films demonstrated a record low Gilbert damping coefficient \cite{scheck}. This material is therefore of interest for spin transfer devices. However, such devices are generally composed of polycrystalline layers. Therefore it is of interest to examine polycrystalline FeV films to assess their characteristics and device potential.

\begin{figure}
\begin{center}\includegraphics[width=9cm]{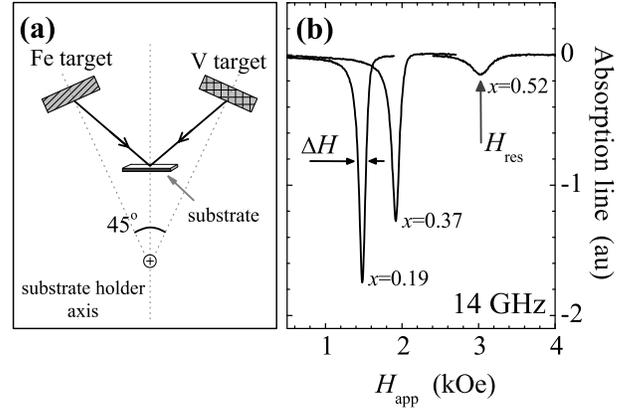}
\vspace{-4 mm}\caption{\label{fig1} a) The co-sputtering setup. b) Typical absorption curves at 14 GHz for a selection of $||$Ta$|$Cu$|$7.5 nm Fe$_{1-x}$V$_x$$|$Cu$|$Ta$||$ films, with $x$=0.19, 0.37 and 0.52. The resonance field $H_{\mathrm{res}}$ and the linewidth $\Delta H$ are indicated.}
\vspace{-6 mm}
\end{center}
\end{figure}

In this paper, we present a FMR study of thin polycrystalline Fe$_{1-x}$V$_x$ alloy films with $0 \leq x < 0.7$ grown by co-sputtering. The FeV layers were embedded between two Ta$|$Cu layers, resulting in the layer structure $||$5 Ta$|$10 Cu$|$FeV$|$5 Cu$|$10 Ta$||$, where the numbers are layer thickness in nm. FeV polycrystalline films were prepared by dc magnetron sputtering at room temperature from two separate sources, oriented at a 45$^{\mathrm{o}}$ angle (Fig. \ref{fig1}a). The substrate, cut from a Silicon wafer with 100 nm thermal oxide, was 64 mm long and about 5 mm wide. The Fe and V deposition rates were found to vary linearly across the wafer. The Fe and V rates were then adjusted to produce a film in which $x$ varies from 0.37 to 0.66 across the long axis of the wafer. The base pressure in the UHV chamber was  $5 \times 10^{-8}$ Torr and the Ar pressure was set to 3.5 mTorr. The FeV was 7.5 nm in thickness, varying by less than 0.3 \% across the substrate. An Fe$_{1-x}$V$_x$ film 3 nm thick was also fabricated. To produce films with $x < 0.30$ the rate of the V source was decreased. Finally, pure Fe films with a thickness gradient ranging from 7 nm to 13.3 nm were deposited.

The FMR measurements were carried out at room temperature using a coplanar wave-guide (CPW) and the flip-chip method. Details of the experimental setup and of the CPW structural characteristics can be found in \cite{beaujour}. A dc magnetic field, up to 10 kOe, was applied in the film plane, perpendicular to the ac magnetic field. Absorption lines at frequencies from 2 to 50 GHz were measured by monitoring the relative change in the transmitted signal as a function of the applied magnetic field.

\section{Results}

\begin{figure}
\begin{center}\includegraphics[width=9cm]{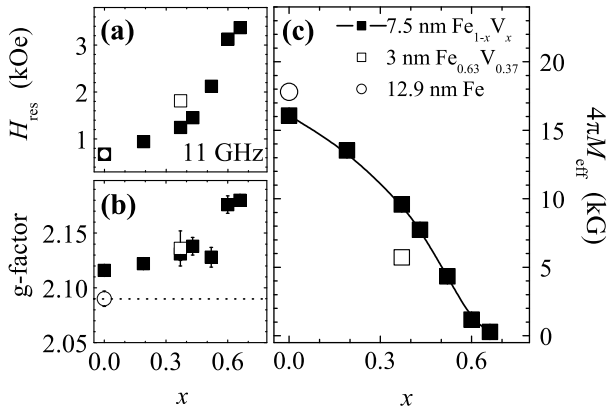}
\vspace{-8mm}\caption{\label{fig2} a) The resonance field at 11 GHz versus $x$ and b) the effective demagnetization field versus $x$. The solid line is a guide to the eye. c) The Land$\acute{e}$ $g$ factor as a function of $x$. The dotted line shows the $g$-factor value of bulk Fe.}
\vspace{-6 mm}
\end{center}
\end{figure}

Typical absorption lines at 14 GHz of selected FeV alloy films are shown in Fig. \ref{fig1}b. The lines are lorentzian for most frequencies. At a fixed frequency, the FMR absorption decreases with increasing V content. The FMR peak of a film 7.5 nm thick with $x=0.66$ is about 100 times smaller than that of a pure Fe of the same thickness. This is accompanied by a shift of $H_{\mathrm{res}}$ towards higher field values (Fig. \ref{fig2}a). The effective demagnetization field $4 \pi M_{\mathrm{eff}}$ and the Land$\acute{e}$ g-factor $g$ were determined by fitting the frequency dependence of the resonance field $H_{\mathrm{res}}$ to the Kittel formula \cite{kittel2}:
\begin{equation}
\label{e.0} f^2=\left(\frac{g \mu_{\mathrm{B}}}{\mathrm{h}}\right)^2 H_{\mathrm{res}}(H_{\mathrm{res}}+4 \pi M_{\mathrm{eff}}),
\end{equation}
where the effective demagnetization field is:
\begin{equation}
\label{e.01} 4 \pi M_{\mathrm{eff}}=4 \pi M_{\mathrm{s}}-H_{\perp}.
\end{equation}
Note that in the absence of a perpendicular anisotropy field $H{_{\perp}}$, the effective field would be directly related to $M_{\mathrm{s}}$.
The dependence of $4 \pi M_{\mathrm{eff}}$ on V concentration is shown in Fig. \ref{fig2}c. As $x$ increases the effective demagnetization field decreases dramatically, going from about 16 kG for $x=0$ to 1.1 kG for $x=0.66$. Note that the effective demagnetization field of the 7.5 nm Fe film is about 25 \% lower than that of bulk Fe (21.5 kG). The 12.9 nm Fe film exhibits a larger $4 \pi M_{\mathrm{eff}}$, which is, however, still lower than $4 \pi M_{\mathrm{s}}$ of the bulk material. Similarly, the  $4 \pi M_{\mathrm{eff}}$ of an Fe$_{0.63}$V$_{0.37}$ film is thickness dependent: decreasing with decreasing layer thickness.

The Land$\acute{e}$ g-factor increases monotonically with increasing V concentration (Fig.  \ref{fig2}b). The minimum $g$-factor is measured for the Fe film: $g=2.11 \pm 0.01$, which is slightly larger than the value of bulk material ($g=2.09$). Note that $g$ of a Fe film 12.9 nm thick is equal to that of Fe bulk. However, the $g$-factor of the Fe$_{0.63}$V$_{0.37}$, does not appear to be thickness dependent: the 3 nm Fe$_{0.63}$V$_{0.37}$ layer has about the same $g$ value than the 7.5 nm Fe$_{0.63}$V$_{0.37}$ layer.

\begin{figure}
\begin{center}\includegraphics[width=8cm]{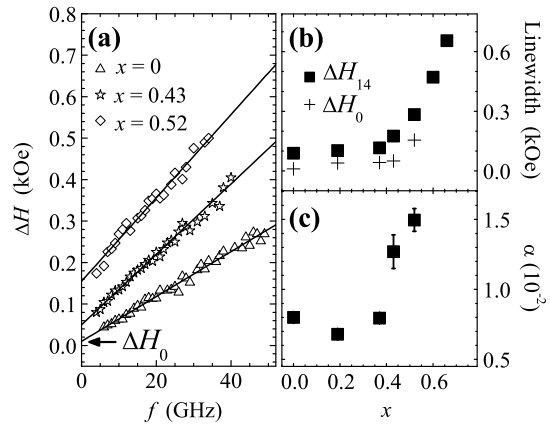}
\vspace{-3 mm}\caption{\label{fig3}a)Frequency dependence of the linewidth for 7.5 nm Fe$_{1-x}$V$_x$ alloy film with $x$=0, 0.43 and 0.52. The solid lines are the best linear fit of the experimental data. b) $\Delta H_{14}$ , the linewidth at 14 GHz , and $\Delta H_0$ are shown as a function of $x$. c) The magnetic damping parameter versus V concentration.}
\vspace{-6 mm}
\end{center}
\end{figure}

The half-power linewidth, $\Delta H$, was studied as a function of the frequency and of the V concentration. Fig. \ref{fig3}b shows the dependence of the FMR linewidth on $x$ at 14 GHz. The general trend is that $\Delta H$ increases with $x$. However, there are two regimes. For $x>0.4$, the linewidth depends strongly on $x$, increasing by a factor 5 when $x$ is increased from 0.4 to 0.66. The dependence of the linewidth on $x$ is more moderate  for the films with $x<0.4$: it increases by about 30 \%. For all samples, the linewidth scales linearly with the frequency. A least square fit of $\Delta H (f)$ gives $\Delta H_0$, the intercept at zero frequency, and the Gilbert damping parameter $\alpha$ which is proportional to the slope: $d \Delta H/df=(2 \mathrm{h}/g \mu_{\mathrm{B}}) \alpha $ \cite{mills}. $\Delta H_0$ is typically associated with an extrinsic contribution to the linewidth and related to magnetic and structural inhomogeneities in the layer. For two samples with the highest Vanadium concentration, $x= 0.60$, 0.66, the linewith is dominated by inhomogeneous broadening and it was not possible to extract $\alpha$. As $x$ increases, $\Delta H_0$ and $\alpha$ increases. The damping parameter and $\Delta H_0$ remain practically unchanged for $x \leq 0.4$ and when $x>0.4$, both the intercept and the slope of $\Delta H$ versus $f$ increase rapidly.

\section{Discussion}
Several factors can contribute to the dependence of $4 \pi M_{\mathrm{eff}}$ on the V concentration. First, the decrease of the effective demagnetization field can be associated with the reduction of the alloy magnetization density $M_{\mathrm{s}}$ since the Fe content is reduced. In addition, a neutron scattering study showed that V acquires a magnetic moment antiparallel to the Fe, and that the Fe atom moment decreases with increasing V concentration \cite{mirebeau}. The Curie temperature of Fe$_{1-x}$V$_x$ depends on $x$. In fact, $T_{\mathrm{c}}$ for $x$=0.65 is near room temperature \cite{kakehashi}. It is important to mention that a factor that can further decrease $4 \pi M_{\mathrm{eff}}$ is an out-of-plane uniaxial anisotropy field $H_{\perp}$ (Eq. \ref{e.01}). In thin films, the perpendicular anisotropy field is commonly expressed as $H_{\perp}=2 K_{\perp}/(M_{\mathrm{s}} t)$, where $K_{\perp}>0$ is the anisotropy constant and $t$ the ferromagnetic film thickness \cite{kim}. In this simple picture, it is assumed that $K_{\perp}$ is nearly constant over the thickness range of our films. This anisotropy can be associated with strain due to the lattice mismatch between the FeV alloy and the adjacent Cu layers and/or with an interface contribution to the magnetic anisotropy.  
For Fe films with $t=7.5$ and 12.9 nm, a linear fit of 4$\pi M_{\mathrm{eff}}$ versus $1/t$ gives $4 \pi M_{\mathrm{s}}=20.2$ kG and $K_{\perp}=2.5$ erg/cm$^2$. The value extracted for $4 \pi M_{\mathrm{s}}$ is in the range of the value of Fe bulk. A similar analysis conducted on Fe$_{0.63}$V$_{0.37}$ films of thickness $t$=3 and 7.5 nm gives 4$\pi M_{\mathrm{s}}=12.2$ kG and $K_{\perp}= 0.1$ erg/cm$^2$. The result suggests that the surface anisotropy constant decreases with increasing $x$.
\section{Summary}
The effective demagnetization field of the polycrystalline Fe$_{1-x}$V$_{x}$ alloy films decreases with increasing $x$ and almost vanishes for $x \approx 0.7$. A FMR study on epitaxial films have shown a similar $x$ dependence of $4 \pi M_\mathrm{eff}$ \cite{scheck}. 
Using the value of $M_{\mathrm{s}}$ calculated in the analysis above, we estimate the Gilbert damping constant of a 7.5 nm Fe layer and 7.5 nm Fe$_{0.63}$V$_{0.37}$ alloy film to be $\mathrm{G}_{\mathrm{Fe}}=239$ MHz and $\mathrm{G}_{\mathrm{FeV}}=145$ MHz respectively. The decrease of the effective demagnetization field of Fe$_{1-x}$V$_x$ with increasing $x$ is accompanied by a decrease of the Gilbert damping constant. A similar $x$ dependence of $\mathrm{G}$ was observed in epitaxial films \cite{scheck}. The authors explained the decrease of $\mathrm{G}$ by the reduced influence of spin-orbit coupling in lighter ferromagnets. Note that the Gilbert damping of our films is larger than what was found for the epitaxial films (G=57 MHz for epitaxial Fe 8 nm thick).

We note that the Fe$_{0.63}$V$_{0.37}$ alloy film, which has $4 \pi M_{\mathrm{s}}$ approximatly the same as that of Permalloy, has a magnetic damping constant of the same order than that of Py layer in a similar layer structure \cite{mizukami}. Hence, with their low $M_{\mathrm{s}}$ and $\alpha$, polycrystalline FeV alloy films are promising materials to be integrated in spin-tranfer devices.

\end{document}